\documentclass{article}

\usepackage{other/arxiv}
\usepackage{graphicx}
\usepackage{subcaption}

\usepackage[utf8]{inputenc} 
\usepackage[T1]{fontenc}    
\usepackage{hyperref}       
\usepackage{url}            
\usepackage{booktabs}       
\usepackage{amsfonts}       
\usepackage{nicefrac}       
\usepackage{microtype}      
\usepackage{lipsum}

\title{Scalable Data Classification \\for Security and Privacy}

 \author{
   Paulo Tanaka  \\
   Facebook\\
   Menlo Park, CA\\
   \texttt{paulot@fb.com} \\
   \And
   Sameet Sapra \\
   Facebook\\
   Menlo Park, CA \\
   \texttt{ssapra@fb.com} \\
    \And
   Nikolay Laptev \\
   Facebook\\
   Menlo Park, CA \\
   \texttt{nlaptev@fb.com} \\
 }

\begin{document}
\maketitle

\begin{abstract}

Content based data classification is an open challenge. Traditional Data Loss Prevention (DLP)-like systems solve this problem by fingerprinting the data in question and monitoring endpoints for the fingerprinted data. With a large number of constantly changing data assets in Facebook, this approach is both not scalable and ineffective in discovering what data is where. This paper is about an end-to-end system built to detect sensitive semantic types within Facebook at scale and enforce data retention and access controls automatically.

The approach described here is our first end-to-end privacy system that attempts to solve this problem by incorporating data signals, machine learning, and traditional fingerprinting techniques to map out and classify all data within Facebook. The described system is in production achieving a 0.9+ average F2 scores across various privacy classes while handling a large number of data assets across dozens of data stores.

\end{abstract}


\section{Introduction}

Organizations today collect and store large amounts of data in various formats and locations \cite{privacy_ent}. The data is then consumed in many places, sometimes copied or cached several times, causing valuable and sensitive business information to be scattered across many enterprise data stores. When an organization is required to meet certain legal or regulatory requirements, for instance to comply with regulations during civil litigation, it becomes necessary to gather the locations of the data in question. For example, if a privacy regulation states that an organization must mask all Social Security numbers (SSN) when delivering personal information to unauthorized entities, a natural first step is to find all occurrences of SSN within the entire organization’s data stores. In these circumstances, data classification becomes crucial \cite{privacy_ent}. A classification system would enable organizations to automatically enforce privacy and security related policies, such as enabling data retention access control policies. This paper introduces a system that we have built at Facebook that utilizes multiple data signals, scalable system architecture, and machine learning to detect sensitive semantic types within Facebook at scale.

Data discovery and classification is about finding and marking enterprise data in a way that enables quick and efficient retrieval of the relevant information when needed. The current process is rather manual and consists in examining the relevant laws or regulations, identifying which types of information should be considered sensitive and what are the different sensitivity levels, and then building the classes and classification policy accordingly \cite{privacy_ent}. Then, Data Loss Protection (DLP)-like systems fingerprint the data in question and monitor endpoints downstream for the fingerprinted data. When dealing with a data warehouse with a large number of assets and petabytes of data, this approach simply does not scale.

Our objective is to build a data classification system that scales to both persistent and non-persistent user data, with no additional constraints on the type or format of data. This is a bold goal, and naturally comes with challenges. Any data record can be thousands of characters long, so we must efficiently represent it using a common set of features that can later be aggregated and moved around easily. These features must not only enable accurate classification, but also provide the flexibility and extensibility to easily add detection for new data types in the future. Second, we must deal with large offline tables. Persistent data may be stored in tables that are many petabytes in size, potentially resulting in slow scanning speeds. Third, we must comply to tight classification SLA’s on non-persistent data. This forces our classification system to be highly performant and provide classification results accurately and efficiently. Finally, we must ensure a low latency data classification for our non-persistent data so that we can also perform data classification in real time as well for online use cases.

In this paper we describe how we dealt with the above challenges and introduce a fast and scalable classification system that classifies data elements from all types, formats, and sources based on a common set of features. We have scaled up our system architecture and built a dedicated machine learning model to quickly classify both offline and online data at Facebook. This paper is organized as follows: In Section \ref{sys_arch} we introduce the overall system design followed by Section \ref{ml_arch} with discussion about the machine learning part of the system. In Sections \ref{related_work} and \ref{conclusion} we, respectively, introduce related work and outline our future direction.

\section{System Architecture}
\label{sys_arch}

To deal with the challenges of both persisted data and online data at the scale of Facebook, our data classification system has two separate flows that we will discuss at length.

\begin{figure}
\centering
\includegraphics[width=16cm]{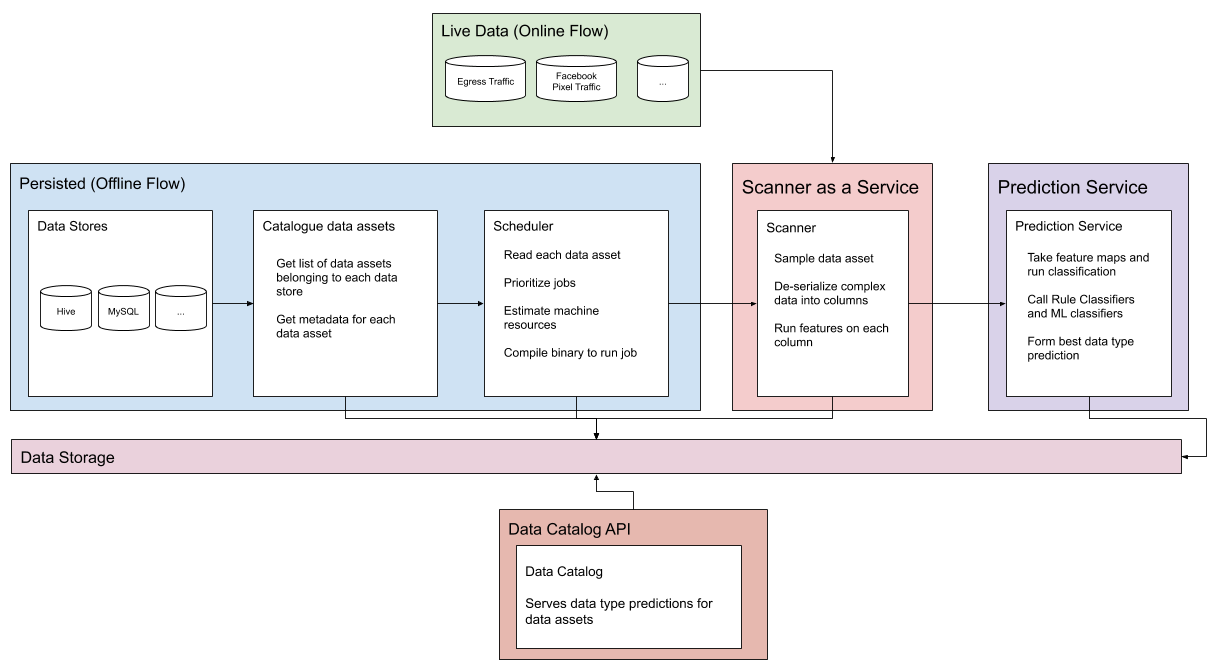}
\caption{Online and offline prediction flows}
\label{fig:arch2}
\end{figure}

\subsection{Persisted Data}

Initially, our system must learn about the universe of data assets at Facebook. For each data store, we collect some basic information such as the datacenter containing that data, the system containing that data, and the assets located in a given data store. This forms a metadata catalog that enables the system to efficiently retrieve data without overloading other clients and resources used by other engineers.

This metadata catalog provides a source of truth for all assets to scan and enables us to keep track of the state of different assets. With this information, we establish a scheduling priority based on this collected data and internal information from the system such as the last time an asset was successfully scanned and how recently the asset was created, and past memory and CPU requirements for that asset if it was scanned before. Then, for each data asset, as resources become available, we invoke a job to perform the actual scanning of the data asset.

Each of these scanning jobs is a compiled binary that performs a Bernoulli sample on the latest data available for each asset. The asset is broken down into individual columns, where each column’s classification result is treated independently. Additionally, the system scans any rich data inside columns, including json, arrays, encoded structs, URLs, base 64 serialized data, and more. This can significantly increase the running-time for scanning a data asset, as a single table may have thousands of nested columns in a json blob.

For each row that was sampled in the data asset, the classification system extracts float and text features from the content and associates each feature back to the column it was taken from. Thus, the output of the feature extraction step is a map of all features for each column that the system found in the data asset.

\subsubsection{Why features?}

The notion of features is a pivotal moment for our system. Instead of float and text features, we could be passing around raw string samples that were directly fetched from each data asset. Furthermore, our machine learning models could be trained on each sample directly, rather than hundreds of feature counts that only attempt to approximate the sample.

There are a few key reasons for this:

\begin{enumerate}
\item Privacy-first: Most importantly, the notion of features allows us to only store the samples we fetch in memory. This ensures we only store the samples for a singular purpose and don’t log them ever in our own efforts to classify the data. This is particularly important with non-persisted data as the service must maintain some classification state before providing a prediction.
\item Memory: Some samples can be thousands of characters long. Storing this data and passing it to parts of our system needlessly consumes many extra bytes. This can compound over time, given that there are many data assets with thousands of columns.
\item Aggregating features: With features, we can cleanly represent the results of each scan through that feature set, which allows the system to merge results from previous scans of the same data asset in a convenient way. This can be useful for aggregating scan results for a single data asset across multiple runs.
\end{enumerate}

These features are then sent to a prediction service, where we employ rule based classification and machine learning to predict data label(s) for each column. The prediction service relies on both rule classifiers and machine learning and chooses the best prediction given from each prediction entity.

The rule classifiers are manual heuristics and use counts and ratios to normalize a feature against a range of 0 to 100. Once this initial score is generated for each data type and the column name associated with this data does not fall in any “deny lists”, the rule classifier picks the highest normalized score among all data types.

Due to the classification complexity involved, using manual heuristics exclusively results in a sub-par classification accuracy, especially for unstructured data. For this reason, we have also developed a machine learning system to deal with classification of unstructured data, like user generated content and address. Using machine learning enabled us to begin moving away from manual heuristics and to leverage additional data signals (e.g., column names, data lineage) greatly improving detection accuracy. We will deep dive into our machine learning architecture later on in the paper.

The prediction service stores the results for each column, along with metadata regarding the scan time and state. Any consumers and downstream processes that depend on this data can read it from a daily published dataset that aggregates the results of all these scan jobs or a real-time API from our data catalog. The published predictions serve as the backbone for automatic privacy and security policy enforcement.

Finally, after the prediction service writes all data and all predictions are stored, our data catalog API can return all the data type predictions for a given data asset in real-time. Each day, the system also publishes a dataset containing all the latest predictions for each asset that serves as the backbone for automatic policy enforcement.

\subsection{Non-Persisted Data}

While the above process happens for persisted assets, non-persisted traffic is also considered to be part of an organization’s data and can be sensitive as well. For this reason, our system provides an online API for generating real-time classification predictions for any non-persisted traffic. This live prediction system is heavily used to classify egress traffic, incoming traffic to machine learning models, and advertiser data that flows through our systems.

Here, our API takes in two main arguments: a grouping key and the raw data to be predicted. The service will perform the same feature extraction as described above and group features together for the same key. These features are maintained in a persisted cache as well for failover recovery. For each grouping key, the service ensures it has seen enough samples before invoking the prediction service, following the same process outlined above.

\subsection{Optimizations}

For scanning some data stores, we use libraries and techniques that optimize reads from warm storage \cite{warm_blob_storage} and ensures that there is no disruption from other users accessing that same data store.

For extremely large tables (50+ petabytes), despite all the optimizations and memory efficiencies employed, the system can still struggle to scan and compute everything before running out of memory. After all, the scan is computed entirely in memory and is not persisted for its duration. If these large tables have thousands of columns with unstructured blobs of data, the job can fail from lack of memory resources while making predictions for the entire table, decreasing our coverage. To combat this, we’ve optimized our system to use the scan speed as a proxy for how well the system can handle its current load. We use this as a predictive mechanism to predict memory issues and pre-emptively calculate the feature map using less data than normal.

\subsection{Data Signals}

A data classification system is only as good as the signals that are provided. In this section we go over all the signals leveraged by the classification system.

\begin{itemize}

\item Content Based: Of course, the first signal of most importance in this system is the content. We perform a Bernoulli sample over each data asset we scan and run our feature extraction over the content of the data itself. Many of the features in our system are derived from the content. We can have any number of float features that represent counts for how many times a certain type of sample was seen. For example, we can have features for the number of emails seen in the sampling, or a feature for how many emojis were seen in the samples. These feature counts can be normalized and aggregated across different scans.

\item Data Lineage: Lineage is an important signal that can help us when the content has changed from its parent table. A common example here is hashed data. When the data in a child table is hashed, this hashed data often can come from a parent table, where the data remains in plaintext. Lineage data can help the system classify certain types of data when the data is not clearly readable or has been transformed from a table upstream.

\item Annotations: Annotations are another high quality signal that helps us identify unstructured data. In fact, annotations and lineage data can work together to propagate attributes across different data assets. Annotations help us identify the source of unstructured data while lineage data can help us track the flow of this data throughout the data warehouse.

\item Data Injection: Lastly, data injection is a technique where we intentionally inject special, unreadable characters into known sources that contain known data types. Then, whenever we scan content that contains that same, unreadable sequence of characters, we can deduce that the content must have been derived from that known data type. This is another extremely high data signal similar to annotations except that content-based detection can help us detect injected data.

\end{itemize}

\subsection{Measuring Metrics}

A significant component of a classification system is having a rigorous methodology for measuring metrics. In a classification system, the primary metrics for iterating on classification improvements is the precision and recall for each label, with the F2 score as the system’s topline metric.

To calculate these metrics, the system needs an independent methodology for labeling data assets that does not rely on the system itself but can be used to compare directly to the system. In this section, we describe how we collect ground truth from within Facebook and use it to train our classification system.

\subsubsection{Ground Truth Collection}

We accumulate ground truth from each source listed below into its own table, where each table is responsible for aggregating the latest ground truth from that specific source. Each source has data quality checks to ensure that the ground truth for each source is high quality and contains the latest data type labels.

\begin{itemize}

\item Logging framework configs: Certain fields in Hive tables are filled in with data that is known to be of a certain data type. Leveraging and propagating this data serves as a reliable, high throughput ground truth.
\item Manual labeling: Developers supporting the system, as well as external labelers, are trained to label columns. This generally works well for all sorts of data types that are present in the data warehouse and can be the primary source of ground truth for some unstructured data, like message data or user generated content.
\item Lineage column propagation: Columns from parent tables can be tagged or annotated to contain certain data, and we can track this data in downstream tables.
\item Sampling code paths: Code paths within Facebook are known to carry data of a certain type. Using our scanner as a service architecture, we can sample these code paths that have known data types and send it through the system, which already promises not to store this data.
\item Sampling tables: Large Hive tables that are known to contain the entire corpus of data can also be used as training data and sent through the system’s scanner as a service. This is great for tables that contain the entire full range of the data type so that sampling this column randomly is equivalent to sampling the entire universe of that data type.
\item Synthetic data: We can even use libraries that generate data on the fly. This works well for simple, public data types, like address or GPS.
\item Data Stewards: Privacy programs tend to leverage data stewards to manually attach policies to pieces of data. This serves as a high fidelity ground truth to our system.

\end{itemize}

We combine each ground truth source into one corpus containing all the ground truth our system possesses. The biggest challenge with ground truth is to ensure that it’s representative of the data warehouse. Otherwise, classification engines can overfit on the ground truth. To combat this, all the sources above are utilized to ensure a balance when training models or computing metrics. Also, manual labelers uniformly sample different columns in the data warehouse and label the data accordingly so that the ground truth collection remains unbiased.

\subsubsection{Continuous Integration}

To enable rapid iteration and improvement, it’s important to always measure the performance of the system in real-time. We can measure any and all classification improvements against the system today so we can be data-driven and tactical in  further improvements. Here, we will go over how the system completes the feedback loop provided by the ground truth data collection.

When our scheduling system encounters a data asset that has a data type label from a ground truth source, we schedule 2 jobs: one job using our production scanner and thus, our production features, and a release candidate job (RC) job using the latest built scanner and thus, our latest RC features. Each scanner job writes its output to a table, tagging the version of each scanner job along with the classification results.

Thus, we can compare the RC classification results against production classification results in real-time and have a clear way to measure improvements over time.

While our datasets compare RC and PROD features, we also log many variations of the prediction service’s ML classification engine: the most recently built machine learning model, the current model in production, and any experimental models that we are currently testing for potential improvements. The same approach described above allows us to “cut” against different model versions (agnostic of our rule classifiers) and compare metrics in real-time. This makes it easy to determine when a machine learning model experiment is ready to be promoted to production.

Every night, the RC features calculated for that day is sent to the ML model training pipeline, where the ML model trains on the latest RC features and evaluates its performance against the ground truth dataset.

Each morning, the ML model finishes training and publishes the model automatically as an experimental model, where it is fetched and automatically incorporated into the list of experimental models.

\begin{figure}
\centering
\includegraphics[width=16cm]{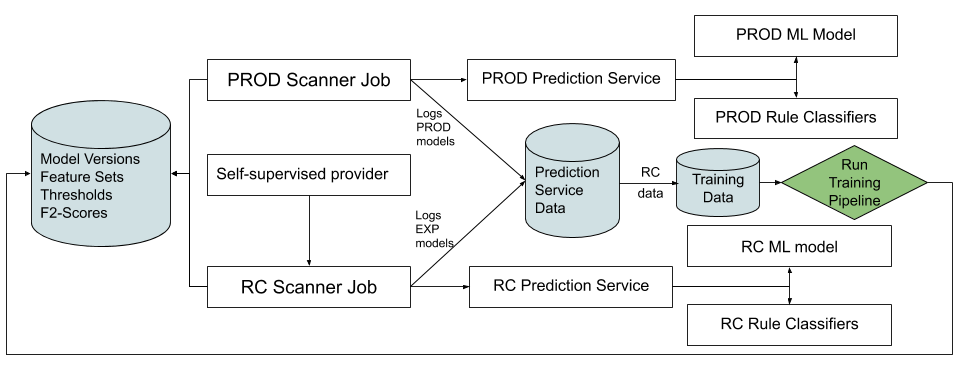}
\caption{The diagram that describes the continuous integration flow for how RC features are generated and sent to the ML model.}
\label{fig:arc1}
\end{figure}

\subsubsection{Some Results}

The system labels over 100 different data types with high accuracy. Data types that are well-structured, like emails and phone numbers, are classified with a f2-score of over 0.95. Data types that are free-formed in content, such as user generated content and name, also perform very well, with f2-scores of over 0.85.

Our system classifies a large number individual columns of both persisted and online data daily across all data stores on a daily basis. Over 500 Terabytes of data are scanned daily in over 10 data stores. The coverage of most of these data stores is over 98

The system has become very efficient over time, with classification jobs in the persisted, offline flow taking 35 seconds on average to complete from scanning the asset to computing predictions for each column.

\section{Machine Learning System Component}
\label{ml_arch}

In the previous section, we did a deep dive into the architecture of the overall system, highlighting the scale of the system, the optimizations, and the offline and online data flows. In this section, we will zoom in on the prediction service and describe the machine learning system that powers the prediction service.

\begin{figure}
\centering
\includegraphics[width=12cm]{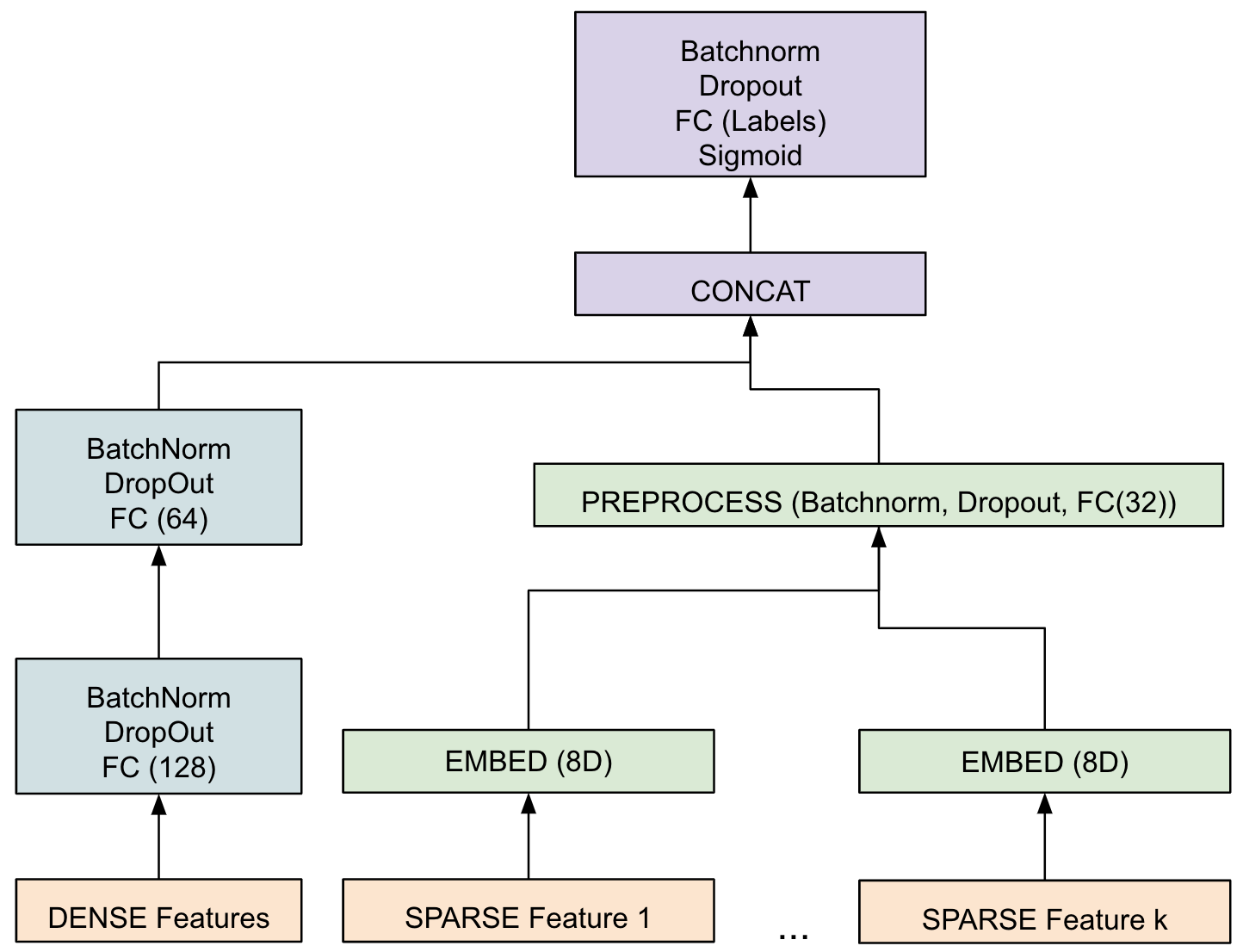}
\caption{A high level diagram of the Machine Learning component of our data classification system.}
\label{fig:ml_model}
\end{figure}

With over 100 data types and some unstructured content like message data and user generated content, using manual heuristics exclusively results in a sub-par classification accuracy, especially for unstructured data. For this reason, we have also developed a machine learning system to deal with the complexities of unstructured data. Using machine learning has enabled us to begin moving away from manual heuristics and to leverage features and additional data signals (e.g., column names, data lineage) to improve detection accuracy.

The implemented model learns embeddings \cite{NIPS2013_5021} over dense and sparse features separately which are then concatenated to form a vector that is passed through a series of batch normalization \cite{pmlr-v37-ioffe15} stages and non-linearities to produce a final output. The final output is a float between [0-1] for each label indicating a probability of how likely a given example to belong to a given sensitivity type. Leveraging PyTorch for the model allowed us to move faster, providing the ability for others outside the team to quickly make and test changes.

During architecture design it was critical to model sparse (e.g., text-based) and dense (e.g., numeric) features separately due to their intrinsic difference. For the final architecture, it was also critical to do a parameter sweep in order to find optimal value for the learning rate, batch size and other model hyper-parameters. The choice of the optimizer used was also an important hyper-parameter and we found that the popular \textit{Adam} optimizer often results in a model overfit whereas \textit{SGD} often produces a more stable model. There were additional nuances that we had to incorporate directly into the model, such as static rules which ensured that the model outputs a deterministic prediction when a given feature had a particular value. These static rules were specified by our customers and we found that incorporating them directly into the model resulted in a more self-sustained and reliable architecture as opposed to implementing a post-processing step to handle these special edge cases. Also note that during training these rules were turned off to not interfere with gradient descent training process. 

\subsubsection{Challenges}

One of the challenges we faced was collecting high quality ground data. The machine learning model needs ground truth data for each class so that it can learn associations between features and labels. In an earlier section, we discussed our methods of collecting ground truth to both measure the system and to train models. Analysis on that data revealed that data classes like credit card numbers and bank account numbers are not extremely prevalent in our warehouse, making it difficult to gather large amounts of ground truth data to train our models.
To combat this issue, we developed processes to generate synthetic ground truth for these classes. We are generating synthetic ground truth for sensitive data types, including \textit{SSN}, \textit{credit card} numbers and \textit{IBAN} numbers for which the machine learning model could not predict previously. This approach allows us to handle sensitive data types without the privacy risk associated with harboring real sensitive data.

Beyond the challenges with high quality ground data, there are also open architectural challenges that we are working on such as \textit{isolating changes} and \textit{early stopping}. Isolating changes is important so that when various changes are made to different parts of the network, the impact is isolated to specific classes and does not have a wide-spread impact on the overall prediction performance. Improving our early stopping criteria is also critical in order for us to stop the training process at a stable point for \textit{all} the classes as opposed to a point where some classes overfit and some underfit.

\subsubsection{Feature Importance}

When a new feature is introduced into the model, we would also like to know its overall impact to the model. We also would like to ensure that the model predictions are still interpretable by a human so that we can precisely understand which features are being leveraged for each data type. To that end, we have developed a \textit{per-class} feature importance for our PyTorch model. Note that this is different from the overall feature importance that is typically supported because it does not tell us which features are important for a specific class. We measure the importance of a feature by calculating the increase in the model’s prediction error after permuting the feature. A feature is ``important" if shuffling its values increases the model error, because in this case the model relied on the feature for the prediction. A feature is “unimportant” if shuffling its values leaves the model error unchanged, because in this case the model ignored the feature for the prediction \cite{feature_imp_cite}.

Per class feature importance allows us to make the model interpretable so we can see what the model pays attention to when predicting a given label. For example when we would analyze label \textit{ADDR} we would ensure that an address-related feature, such as \textit{AddressLinesCount}, is ranked high on the per-class feature importance table to ensure our human intuition aligns well with what the model learned.

\subsubsection{Evaluation}

It is important to identify a single metric of success. For this purpose we have picked \textit{F2} which provides a balance between recall and precision (biasing recall a bit more). Recall is more important for the privacy use-case than precision because it is critical for the team to ensure not to miss any sensitive data (while ensuring a reasonable precision). Actual evaluation data of the F2 performance of our model is beyond the scope of this paper, nevertheless, with careful tuning we are able to achieve a high (0.9+) F2 score for our most important sensitive classes.

\section{Related Work}
\label{related_work}

There are many existing algorithms for automatic classification of unstructured documents using various techniques such as pattern matching, document similarity search and different machine learning techniques (Bayesian, Decision Trees, k-Nearest Neighbor and more) \cite{Phyu_surveyof}. Any of the above can be used as part of the classification process, the problem however is with scalability. The classification approach described in this paper is biased towards flexibility
and performance allowing us to support new classes in the future and maintain a low latency.

There is a body of work around data fingerprinting as well. For example authors in \cite{7038200} describe a solution that focuses on solving a problem of catching sensitive data leaks. The main assumption provided by the authors is the ability to fingerprint the data in order to match it to a set of known sensitive data. Authors in \cite{appintent} describe a similar privacy leakage problem but their solution is based on a specific Android architecture and only classifies if the user actions led to the private information being sent or if the underlying application leaking user data. The situation presented in this paper is slightly different because user data can also be highly unstructured and therefore we need a more elaborate technique than fingerprinting to use for classification.

Finally, to deal with the lack of data for some sensitive data types, we have introduced synthetic data. There is a large body of literature on data augmentation, for example, authors in \cite{DBLP:journals/corr/abs-1904-12848} investigate the role of noise injection during training and observe positive results for supervised learning. Our approach in privacy differs, because injecting noisy data may be counter-productive, and instead we focus on high quality synthetic data.

\section{Conclusion}
\label{conclusion}

In this paper, we have introduced a system that can classify a piece of data, which allows us to build enforcement systems that ensure that privacy and security policies are followed. We have shown that scalable infrastructure, continuous integration, machine learning, and high quality ground truth data play a key role in the success of many of our privacy initiatives.

There are many axes in which this work can be extended. Future work may include providing support for un-schematized data (files), classifying not only the data type but also the sensitivity level of data, and leveraging self-supervised learning directly during training via generating the exact synthetic examples which will help the model decrease loss by the greatest amount. Future work may also include focus on the investigation workflow where we go beyond detection and provide root cause analysis for various privacy violations, which would help in cases such as sensitivity analysis (i.e., is the privacy sensitivity of a given data type high (e.g. user IP) or low (e.g., an internal Facebook IP)).

\bibliographystyle{unsrt}  
\bibliography{other/references}

\end{document}